\documentclass[twocolumn,pre,showpacs,superscriptaddress]{revtex4}
\usepackage{graphicx}%
\usepackage{psfrag}
\usepackage{dcolumn}
\usepackage{amsmath}
\usepackage{amssymb}
\usepackage{latexsym}
\usepackage{hyperref}
\usepackage{color}
\begin{document}

\def\bea{\begin{eqnarray}}
\def\eea{\end{eqnarray}}
\def\beq{\begin{equation}}
\def\eeq{\end{equation}}
\def\f{\frac}
\def\k{\kappa}
\def\e{\epsilon}
\def\ve{\varepsilon}
\def\be{\beta}
\def\D{\Delta}
\def\h{\theta}
\def\t{\tau}
\def\a{\alpha}

\def\cDa{d\varsigma}
\def\cD{{\cal D}[x]}
\def\cL{{\cal L}}
\def\cLo{{\cal L}_0}
\def\cLa{{\cal L}_1}

\def\Re{{\rm Re}}
\def\sj{\sum_{j=1}^2}
\def\rk{\rho^{ (k) }}
\def\rek{\rho^{ (1) }}
\def\cek{C^{ (1) }}
\def\rz{\rho^{ (0) }}
\def\rt{\rho^{ (2) }}
\def\rtb{\bar \rho^{ (2) }}
\def\trk{\tilde\rho^{ (k) }}
\def\trek{\tilde\rho^{ (1) }}
\def\trz{\tilde\rho^{ (0) }}
\def\trt{\tilde\rho^{ (2) }}
\def\r{\rho}
\def\tD{\tilde {D}}

\def\s{\sigma}
\def\kb{k_B}
\def\la{\langle}
\def\ra{\rangle}
\def\nn{\nonumber}
\def\up{\uparrow}
\def\dn{\downarrow}
\def\S{\Sigma}
\def\dg{\dagger}
\def\d{\delta}
\def\p{\partial}
\def\l{\lambda}
\def\L{\Lambda}
\def\G{\Gamma}
\def\o{\Omega}
\def\w{\omega}
\def\g{\gamma}

\def\noi{\noindent}
\def\a{\alpha}
\def\d{\delta}
\def\p{\partial} 

\def\la{\langle}
\def\ra{\rangle}
\def\e{\epsilon}
\def\n{\eta}
\def\g{\gamma}
\def\break#1{\pagebreak \vspace*{#1}}
\def\hf{\frac{1}{2}}
\title{Modified fluctuation-dissipation and Einstein relation at non-equilibrium steady states }
\author{Debasish Chaudhuri}
\email{d.chaudhuri@amolf.nl}
\affiliation{
FOM Institute for Atomic and Molecular Physics,               
Science Park 104,
1098XG Amsterdam, 
The Netherlands
}
\altaffiliation{Current address: Indian Institute of Technology Hyderabad, 
Yeddumailaram 502205,
Andhra Pradesh, India}
\author{Abhishek Chaudhuri}
\email{a.chaudhuri1@physics.ox.ac.uk}
\affiliation{
Department of Biomedical Science, University of Sheffield,
Western Bank, Sheffield S10 2TN, United Kingdom
}
\affiliation{
The Rudolf Peierls Centre for Theoretical Physics,
University of Oxford, 1 Keble Road, Oxford OX1 3NP, United Kingdom
}

%\\

\date{\today}

\begin{abstract}
Starting from
the pioneering work of G. S. Agarwal [{\em Zeitschrift f{\"u}r Physik} {\bf 252}, 25 (1972)], 
we present a unified derivation of a number of
modified fluctuation-dissipation relations (MFDR) that relate response to small perturbations 
around non-equilibrium steady states  to steady-state correlations. Using this formalism 
we show the equivalence of velocity forms
of MFDR derived using {\it continuum} Langevin and {\it discrete} master equation dynamics.
The resulting additive correction to the Einstein relation is exemplified using a flashing 
ratchet model of molecular motors.
\end{abstract}

\maketitle
\section{Introduction}
Derived within linear response theory, the 
fluctuation dissipation theorem (FDT) predicts how 
the response function of a thermodynamic observable
is related to correlation of  thermal fluctuations at  equilibrium.
Let us assume that an equilibrium system described by a Hamiltonian $H$
is perturbed at time $t=t_1$ by an external force $h(t)$.
The FDT predicts a response at a later time $t_2>t_1$~\cite{Marconi2008}
\beq
R^{eq}_A(t_2-t_1) = \f{\d \la A(t_2)\ra}{\d h(t_1)} = \be \f{\p}{\p t_1}\la A(t_2) [-\p_h H(t_1)]_{h=0}\ra_{eq}
\label{fdt-eq}
\eeq
where the correlation is calculated at equilibrium corresponding to temperature $T$ with $\be = 1/ T$.
The differential operator $\p_h$ in the above relation denotes the scalar derivative 
evaluated at time $t_1$. Thus $-\p_h H$ is the displacement conjugate to $h$ with respect to the Hamiltonian.
Throughout this paper we use Boltzman constant $\kb=1$, unless otherwise stated.
Using the Onsager regression hypothesis the FDT can be interpreted as follows -- the decay of a fluctuation is independent of 
how it has been created, under the influence of a small applied force or spontaneously by thermal noise. 
The FDT is violated away from equilibrium regime 
and this violation has been studied in context of glassy systems, granular matter,
sheared fluid, stochastic processes, and biological systems~\cite{Hanggi1982,Martin2001,Crisanti2003,Marconi2008,Speck2009,Prost2009,Seifert2010,Sarracino2010,Verley2011}. 

In a pioneering study back in 1972~\cite{Agarwal1972}, G. S. Agarwal 
obtained a 
modified fluctuation-dissipation relation (MFDR) that related
response functions around non-equilibrium steady states (NESS)
to correlations evaluated at steady state.
For a system evolving with a statistical dynamics characterized by the Fokker-Planck (FP) equation
$\p_t p = \cLo p$, Agarwal showed that a perturbation in the operator 
$\cLo \to \cLo+h(t) \cLa$ leads to a response that can be expressed in terms of
a correlation function evaluated at the unperturbed steady state~\cite{Agarwal1972,Risken1989},
\bea
R_A(t_2-t_1) = \f{\d\la A(t_2)\ra}{\d h(t_1)} = \la A(t_2) M(t_1) \ra
\eea
where the {\em Agarwal term} $M = [\cLa p_s]/p_s$ with
$p_s$ denoting the steady-state probability distribution. 
Throughout this paper by $\la \dots \ra$ we denote a steady-state 
average. %, unless otherwise stated.

Over the last decade a formalism of stochastic thermodynamics
has been developed that allows %stochastic 
description of  energy and entropy along fluctutating trajectories~\cite{Sekimoto1998,Bustamante2005,Seifert2008}. 
Various fluctuation theorems involving the distribution of 
entropy~\cite{Evans1993,Gallavotti1995,Lebowitz1999,Hatano2001,Evans2002,Esposito2007,Esposito2010a}, 
and work theorems~\cite{Jarzynski1997,Crooks2000,Kawai2007} were discovered.
Recently, using an integral fluctuation theorem, 
%for stochastic non-equilibrium dynamics 
a number of these relations were derived in a
unified manner~\cite{Seifert2005,Seifert2008}. 
Important experimental tests include 
colloidal particles manipulated by laser traps~\cite{Wang2002,Blickle2006,Andrieux2007a},
biomolecules pulled by AFM or laser tweezer~\cite{Liphardt2002,Collin2005} and autonomous motion
of motor proteins~\cite{Hayashi2010}.
Stochastic thermodynamics has also been used  to derive several versions of MFDR
around NESS~\cite{Cugliandolo1994,Lippiello2005,Speck2006,Chetrite2008,Prost2009,Baiesi2009,Seifert2010,Verley2011}. 
Some of these 
predictions were experimentally verified~\cite{Blickle2006, Gomez-Solano2009}.

In this paper, we present a unified derivation of a number of MFDRs based entirely on the
Agarwal formalism~\cite{Agarwal1972}. 
Thus the MFDRs we obtain are intrinsically equivalent to each other.
We show that the Agarwal term $M$ can be expressed as a velocity excess
from a local mean velocity using both the {\em continuum} Langevin and {\em discrete} master equation dynamics.
This interpretation leads us to a modified Einstein relation 
that has the same additive correction term for the two cases. 
Finally we apply this framework to a flashing ratchet model of molecular 
motors~\cite{Julicher1997,Astumian2002,Kolomeisky2007} to calculate the MFDR and the 
additive correction in Einstein relation,
which shows a non-monotonic variation with the asymmetry parameter of the ratchet.

The structure of this paper is as follows. In Sec.~\ref{agarwal} we review the derivation 
of the Agarwal form of MFDR,  
that we use throughout this paper 
to calculate other versions of MFDR expressed in physically observable form. 
Using this result, in Sec.~\ref{entropy-mfdr} we present a simple and straightforward derivation 
of the MFDR in terms of stochastic entropy production, keeping in mind that this relation
was used earlier to derive velocity-MFDR for a master equation dynamics~\cite{Verley2011}.
Then, directly using the Agarwal form, we derive the velocity-MFDR for a 
system evolving with continuum Langevin dynamics in Sec.~\ref{lange-mfdr}, and a discrete
master equation in Sec.~\ref{master-mfdr}. The velocity-MFDR is used in Sec.~\ref{einstein} 
to derive a modified Einstein relation at NESS. 
In Sec.~\ref{ratchet}, we study the 
velocity-MFDR, and the violation of the Einstein relation in a flashing ratchet model of molecular motors. 
Finally in Sec.~\ref{summary}
we summarise our main results and conclude.

\section{The Agarwal form of MFDR}
\label{agarwal}
The probability distribution $p(\varsigma,t)$  of finding a system at state $\varsigma$ 
%(denoting a phase-space coordinate)
at time $t$ evolves with time as
\bea
\p_t p(\varsigma,t) = \cL(\varsigma,h) p(\varsigma,t) % \nn\\
\label{eq1}
\eea
where  $\cL$ is a general time evolution operator that depends on external force $h(t)$. 
For weak $h$, Taylor expanding the operator we get %the form
\bea
\cL(\varsigma,h) %= \cL(0) + h  \p_h \cL 
= \cLo(\varsigma) + h(t) \cLa(\varsigma)
\label{cL}
\eea
where $\cLa = [\p_h \cL]_{h=0}$.
The solution to Eq.~\ref{eq1} is
\beq
p(\varsigma,t) = p_s + \int_{-\infty}^t d\t e^{\cLo (t-\t)} h(\t) \cLa p_s(\varsigma)
\label{ps}
\eeq
where $p_s$ denotes the steady-state distribution that obeys $\cLo p_s=0$.
Then the response of any observable $\la A(t) \ra = \int \cDa A(\varsigma) p(\varsigma,t)$ to a force
$h(t)$ is 
\bea
R_A(t_2-t_1) &=& \f{\d \la A(t_2)\ra}{\d h(t_1)} = \int \cDa A(\varsigma) \f{\d p(\varsigma,t_2)}{\d h(t_1)} \nn\\
&=& \int \cDa A(\varsigma) e^{\cLo (t_2-t_1)} \cLa p_s(\varsigma) \nn\\
&=& \int \cDa A(\varsigma) e^{\cLo (t_2-t_1)} M(\varsigma) p_s(\varsigma) \nn\\
\label{resp}
\eea
where in the last step we used the Agarwal term $M(\varsigma) \equiv [\cLa p_s]/p_s$. 
By definition, the  two-time correlation function is 
$\la A(t) B(0) \ra 
= \int d\varsigma \int d\varsigma' A(\varsigma) B(\varsigma') p_2(\varsigma, t; \varsigma',0)$, 
where $p_2(\varsigma, t; \varsigma',0)$ is the joint probability distribution of finding the system
at state $\varsigma'$ at time $0$ and at state $\varsigma$ at time $t$. One can express
$p_2(\varsigma, t; \varsigma',0) = w(\varsigma, t | \varsigma',0)\, p(\varsigma',0)$ where 
$w(\varsigma, t | \varsigma',0)$ is the transition probability. The time evolution 
$\p_t p = \cLo p$ can be solved to obtain the  transition probability at steady state
$w(\varsigma, t | \varsigma',0)= \exp(\cLo t) \d(\varsigma-\varsigma')$. Thus the two-time correlation
at steady state takes the form
$\la A(t) B(0) \ra = \int d\varsigma A(\varsigma) \exp(\cLo t) B(\varsigma) p_s(\varsigma)$.
Therefore we can write Eq.~\ref{resp} as
\beq
R_A(t_2-t_1) 
= \la A(t_2) M(t_1) \ra. 
\eeq
This is the 
{\em Agarwal form} of MFDR~\cite{Agarwal1972}. % (with $M(x,t_1) \equiv [\cLa p_s(x,t_1)]/p_s(x,t_1)$).
The derivation presented here used a continuum notation of the phase space variable $\varsigma$.
However, the result is general, and can be  derived similarly for a system that evolves through
transitions between discrete states (see Eq.~\ref{master-Ag}).

The Agarwal term in its operator form $M(\varsigma) \equiv [\cLa p_s]/p_s$
requires detailed knowledge of the probability distribution at 
steady state. 
In the rest of this paper we focus on expressing this term in physically observable form.

\section{MFDR in terms of stochastic entropy}
\label{entropy-mfdr}
The definition of non-equilibrium Gibb's entropy 
$S = -\int \cDa \,p(\varsigma,t)\,\ln p(\varsigma,t) \equiv \la s(t) \ra$ has recently been used
to get a definition of the stochastic entropy %of a fluctuating trajectory 
$s(t) = -\ln p(\varsigma,t)$~\cite{Seifert2005}. 
For a master equation based discrete dynamics between states denoted by $n(t)$, the stochastic entropy
can be written as $s(t) = -\ln p_{n(t)}$. 
Using this definition we obtain a simple interpretation of the 
Agarwal term in terms of stochastic entropy 
\bea
M &=& \f{1}{p_s} \cLa p_s = \left. \f{\p_h \cL(h) p}{p} \right|_{h=0} \nn\\
&=& \left. \f{\p_h \p_t p}{p}\right|_{h=0} = - \p_t [\p_h s]_{h=0}.
\eea
In deriving the above relation we assumed that $\cL(h)$ is linear in $h$.
We also used the fact that the steady state distribution $p_s = p|_{h=0}$.
Thus $M$ is expressed as time-evolution of a variable 
conjugate to the external force $h$ with
respect to the stochastic system-entropy $s$. In this sense, %stochastic entropy
$s$ in NESS plays the role similar to the Hamiltonian in equilibrium FDT.
We can now write the MFDR at NESS as
\beq
R_A(t_2-t_1)= \f{\p}{\p t_1}\la A(t_2) [-\p_h s(t_1) ]_{h=0} \ra. 
\label{mfdr-s}
\eeq
Ref.~\cite{Seifert2010,Speck2010} found this relation by considering %a special case where 
a perturbation that takes the system %initial steady state 
to a final steady state.
Note that our simple and straightforward derivation does not require  such an assumption, and thus 
the result is more general.

\subsection{Equilibrium FDT} 
The FDT at equilibrium can easily be derived from Eq.~\ref{mfdr-s}. 
If, even in the presence of external perturbation the system remains at equilibrium,
one can write down the probability distributions as
$p = \exp[-\be(H-F)]$
where $F$ is the free energy. This distribution leads to the relation
$[\p_h p]_{h=0} = \be [(\p_h F - \p_h H)p]_{h=0}$. 
Note that  the equilibrium displacement evalutaed at $h=0$ 
is $[\p_h F]_{h=0} = 0$. Thus we get the identity
$[\p_h s]_{h=0} = -[(\p_h p)/p]_{h=0} = \be [\p_h H]_{h=0}$, which leads to the
equilibrium FDT Eq.~\ref{fdt-eq}.

\section{Velocity MFDR  using Langevin equation}
\label{lange-mfdr}
Let us consider
a Langevin system where the dynamics of a particle evolves by 
\beq
v = \mu f + \n
\label{lange}
\eeq
where $v=\dot x$ is the particle velocity, $\mu$ is the mobility, 
and $f$ denotes total force imparted on the particle. The total force $f(x,t)$ consists of
a force due to interaction $F(x)$ and an external time dependent force $h(t)$: $f(x,t)=F(x)+h(t)$. The last term $\n$ 
denotes a thermal noise that obeys $\la \n \ra=0$ and $\la \n(t) \n(0) \ra = 2 D \d(t)$ with $D=\mu  T$, the 
equilibrium Einstein relation. 
The corresponding FP equation is 
\bea
\p_t p(x,t) &=& -\p_x j(x,t) \label{fp}\\
{\rm with,~~~~} j(x,t) &=& (\mu f(x,t)-D\p_x) p(x,t).\nn
\eea

The velocity form of MFDR for a Langevin system was originally derived 
in Ref.~\cite{Speck2006}. 
Here we briefly outline the derivation starting from the 
Agarwal form. Eq. \ref{fp} can be expressed as,
\bea
\p_t p(x,t) = (\cLo + h(t) \cLa) p(x,t), \nn
\eea
where,
$
\cLo= -\p_x (\mu F) + D \p_x^2, 
$
and
$
\cLa = -\mu \p_x. 
$
Thus the Agarwal term %can be written as
$
M = -\mu (\p_x p_s)/p_s, % = \mu \p_x s(x).
$
and 
$T M= -D (\p_x p_s)/p_s$. 
The definition of the steady state current $j_s$ leads to the relation
$D \p_x p_s=\mu F(x) p_s(x) -j_s$. Defining  a local mean velocity 
at steady state $\nu_s(x)=j_s/p_s(x)$
we can then rewrite $TM=-D (\p_x p_s)/p_s= \nu_s(x) - \mu F(x)$.
In this relation, using the Langevin equation at initial steady state ($h=0$), we get
$T M = \nu_s -v + \n$. Thus, the response function
\beq
 T R_A(t_2-t_1)= \la A(t_2) [\nu(t_1) -v(t_1) + \n(t_1)] \ra.
\label{lend}
\eeq
Note that in the Langevin equation $\mu h(t)$ and $\n(t)$ have the same status,
and $A(x,t)$ can be regarded as a functional  
of noise history. 
Then it can be shown that~\cite{Speck2006},
\bea
 T R_A(t_2-t_1)= D\f{\d\la A(t_2)\ra}{\d \n(t_1)} = \hf \la A(t_2) \n(t_1) \ra.
\label{fdt-noise}
\eea
Thus we can write Eq.~\ref{lend} as
\beq
R_A(t_2-t_1)= \be\, \la A(t_2) \left[v(t_1) - \nu(t_1) \right]\ra
\label{lange-vA}
\eeq
This is the {\em velocity form} of MFDR, which for velocity-response gives
\beq
  R_v (t_2-t_1)= \be\, \la v(t_2) [v(t_1)  -  \nu(t_1)] \ra.
\label{fdt-v2}
\eeq
Note that the steady state average of $\nu$ is the same as the mean velocity:
\bea
\la \nu_s \ra = \int_{-L/2}^{L/2} dx p_s(x) \nu_s(x) = \mu \la F \ra - D [p_s]_{-L/2}^{L/2} 
= \la v_s \ra. \nn\\
\label{nus_av}
\eea
The boundary term $[p_s]_{-L/2}^{L/2} =0$ either by a periodic boundary condition~\cite{Speck2006}, 
or by taking the boundaries to infinity where the probabilities vanish.
If the system is at equilibrium $\nu =0$, and we get back the well-known
equilibrium response,
\beq
R^{eq}_v (t_2-t_1)=\be\, \la v(t_2) v(t_1) \ra_{eq}. 
\eeq
Therefore the non-equlibrium MFDR Eq.~\ref{fdt-v2} can be viewed as the equilibrium FDT
with an additive correction $-\be \la v(t_2) \nu(t_1)\ra$.

%{\textcolor{red}{
It is interesting to note that using Eq. \ref{fdt-noise}, we can arrive at a
non-equilibrium MFDR first obtained in Ref. \cite{Cugliandolo1994} for 
continuous Langevin dynamics and subsequently shown to be true for discrete spin
variables (as well as, for conserved  and non-conserved order
parameter dynamics) in Ref. \cite{Lippiello2005}. Defining the position 
correlation function $C_x(t_2,t_1) = \langle x(t_2)x(t_1)\rangle$ and
the corresponding response function  $2TR_x(t_2,t_1) = 
\langle x(t_2)\eta(t_1) \rangle$ (using Eq. \ref{fdt-noise}), 
we get the modified MFDR
\beq
\left( \partial_{t_1} -  \partial_{t_2} \right)C_x(t_2,t_1) 
=  2TR_x(t_2,t_1) + A(t_2,t_1)
\eeq
where $A(t_2,t_1) = \langle \mu f(t_1)x(t_2) - 
\mu f(t_2)x(t_1)\rangle$ is the so-called asymmetry which vanishes in 
the presence of time reversal symmetry. Note that causality demands that
the response of the system at time $t_2$ to a perturbation at time $t_1$,
$R_x(t_2,t_1)$, is nonzero only when $t_2 \ge t_1$.  
Incorporating time translation invariace and time reversal symmetry
restores the equilibrium FDT, 
$TR_x(t_2,t_1) = \partial_{t_1}C_x(t_2,t_1)$.
Also note that, the choice of the observable $V$
in Ref. \cite{Baiesi2009} as an 1D coordinate $x$, reduces the
second term on the r.h.s of Eq. 13 in Ref. \cite{Baiesi2009} to 
$\langle (L-L^*)V(s)Q(t) \rangle = \langle 2(j/\rho)\nabla xQ(t) \rangle 
= \langle \nu Q\rangle$. Now setting $Q \equiv v$ (velocity), leads Eq. 13 in 
Ref. \cite{Baiesi2009} to Eq. \ref{lange-vA} in our manuscript, 
the velocity form of MFDR.
%}

\section{Velocity MFDR  using master equation}
\label{master-mfdr}
We now focus on a master equation system
where the time-evolution occurs via transitions between discrete states. 
Following Ref. \cite{Speck2006}, we first derive the discrete form of the Agarwal term $M$.
Our main contribution in this section is to express $M$ as an excess velocity, and thus
arrive at a {\em velocity form} of MFDR, similar to the Langevin system.

We begin by considering a set of discrete states $\{n\}$ and write down the corresponding
master equation for the probability $p_m(t)$ of finding the system in a state
$m$ at time $t$:
\bea
\partial_t p_m(t) &=& \sum_n [w_{nm} p_{n}(t) - w_{mn} p_{m}(t) ] \crcr
& \equiv & \sum_n{{\cal{L}}_{mn} p_n(t)} 
\eea
where, $w_{mn}$ represents transition rate from state $m$ to $n$ and is 
generally dependent on the external force $h$. The time evolution operator
\bea
{\cal{L}}_{mn} = w_{nm} - \delta_{mn}\sum_k{w_{mk}}.
\eea
If the external force $h(t)$ acting on the system is weak, Taylor expanding
about $h=0$, we get
\bea
{\cal L}_{mn}(h) = (\cLo)_{mn} + h(t)(\cLa)_{mn}. \nn
\eea
In this relation
\bea
(\cLa)_{mn} = w_{nm}\alpha_{nm} - \delta_{mn}\sum_k{w_{mk}\alpha_{mk}} \nn
\eea
where $\alpha_{mn} = [\p_h \ln{w_{mn}}]_{h=0}$ gives the relative change of rates. 
Note that the system  is prepared in a NESS at $h = 0$ characterized
by the stationary distribution $(p_n)_s$. 
Then Eq.~\ref{resp} can be expressed in the discrete notation as
\bea
R_A(t_2 - t_1) &=& \sum_{m,n} A_m [ e^{\cLo(t_2 - t_1)} p(t_1)]_{mn} M_n  \nn\\
&=&  \la A(t_2)M(t_1)\ra
\label{master-Ag}
\eea
where the Agarwal term is
\bea
M_m &=& \f{1}{(p_m)_s}\sum_n{(\cLa)_{mn} (p_n)_s} \crcr
&=& \sum_n{\f{(p_n)_s}{(p_m)_s} w_{nm}\alpha_{nm}} - \sum_n{w_{mn}\alpha_{mn}}. 
\label{master-M}
\eea

Now we use the above relation to derive the velocity form of MFDR.
We assume a %physical
displacement $d_{mn}$ associated with each transition from state $m$ to $n$. 
This has the property $d_{mn} = -d_{nm}$ and
gives a definition of velocity $v(t)=\sum_m \delta(t - \tau_m) d_{m-1,m}$~\cite{Seifert2010a}. 
A generalized detailed balance in presence of the external force $h$
\bea
\f{w_{mn}(h)}{w_{nm}(h)} = \f{w_{mn}(0)}{w_{nm}(0)}
\exp [\be\, h\, d_{mn} ]
\label{loc_eq}
\eea
leads to the following useful relation 
\bea
\a_{mn} - \a_{nm} = \be d_{mn}.
\eea
We also utilize the probability current
\bea
J_{mn} = p_m w_{mn} - p_n w_{nm} = -J_{nm}.
\eea
Then from Eq.~\ref{master-M} we find the velocity form of Agwarwal term,
\bea
M_m &=& \sum_{n}{\f{(p_n)_s}{(p_m)_s}w_{nm} \a_{nm}} - \sum_{n}{w_{mn} \a_{mn}}  \nn \\
&=& \sum_{n}{\f{(p_n)_s}{(p_m)_s}w_{nm}( \a_{mn}} + \be d_{nm}) - \sum_{n}{w_{mn} \a _{mn}} \nn \\
& = & \be \sum_{n}{\f{(p_n)_s}{(p_m)_s}w_{nm}d_{nm}} - \sum_n \f{1}{(p_m)_s}(J_{mn})_s \a_{mn} \nn \\ 
&=& \be (v_m - \nu_m),
\label{M_vel}
\eea

\noindent
where 
\bea
v_m =  \sum_{n} {\frac{(p_n)_s}{(p_m)_s}w_{nm}d_{nm}}, \nn \\
\beta \nu_m = \sum_{n} {\frac{(J_{mn})_s}{(p_m)_s}\a_{mn}}.
\label{vm-num}
\eea
These relations lead to the  {\em velocity form} of MFDR 
\bea
R_A(t_2-t_1) &=& \sum_{m,n} A_m [ e^{\cLo(t_2 - t_1)} p(t_1)]_{mn} [\be (v_m - \nu_m)] \nn\\
&=& \be\la A(t_2) [v(t_1) - \nu(t_1)]\ra.
\label{master-vA}
\eea
Note that Eq.~\ref{master-vA} agrees with the results obtained in Ref.s~\cite{Verley2011,Seifert2010}.
In particular, Ref.~\cite{Verley2011} used the MFDR expressed in terms of 
stochastic entropy of the system $s$ (Eq.~\ref{mfdr-s})
to obtain  Eq.~\ref{master-vA}. They used the total stochastic entropy $s_{tot}=s+s_m$ where $s_m$ is the stochastic
entropy of the medium and showed
\bea
\partial_h \dot{s}_m(t) &=& \sum_m \delta(t - \tau_m) d_{m-1,m} \equiv v(t) = \sum_m \delta_{n(t),m} v_m \nn \\
\partial_h \dot{s}_{tot}(t) &=& \nu(t) =  \sum_m \delta_{n(t),m} \nu_m
\eea
where $v_m$ and $\nu_m$ are given by Eq. \ref{vm-num}.

Note the equivalence of Eq.~\ref{master-vA} with Eq.~\ref{lange-vA}. Indeed the analogy 
of $\nu$ described here with the local mean velocity $[j(x,t)/p(x,t)]$ in the Langevin system
becomes even more clear when we compare the steady state average 
$\la \nu_s \ra = \sum_m (p_m)_s \nu_m$  with $\la v_s \ra = \sum_m (p_m)_s v_m$
and find 
\bea
\la \nu_s \ra  = T\sum_{mn} J_{mn} \a_{mn} = \sum_{mn} (p_n)_s w_{nm} d_{nm}= \la v_s \ra. 
\eea
This relation is the same as Eq.~\ref{nus_av} obtained for the
Langevin system.

For a  velocity-response  Eq.~\ref{master-vA} readily leads us to Eq.~\ref{fdt-v2} already obtained 
in the context of Langevin dynamics. This completes one of the main achievements
of this paper~--~ 
the Agarwal formalism leads to the same form of velocity-MFDR for discrete master equation 
and continuum Langevin dynamics.

\section{Einstein relation}
\label{einstein}
Using the velocity MFDR  (Eq.~\ref{fdt-v2}) 
we find the mobility in NESS
\bea
\mu_s = \int_{0}^\infty d\t  R_v (\t) 
= \be \int_{0}^\infty d\t  \la v(\t) [v(0) - \nu(0)] \ra.
\eea
On the other hand, the diffusion constant in an NESS having mean velocity $\la v_s \ra$ is
\beq
D_s = \int_{0}^\infty d\t \la[v(\t) - \la v_s \ra ][v(0)- \la v_s \ra] \ra.
\eeq
Thus the mobility $\mu_s$ and diffusion constant $D_s$ at NESS do not
satisfy the equilibrium Einstein relation, i.e., $ D_s -T\mu_s =I \neq 0$. 
The difference gives us the modification in %the Green-Kubo or 
the Einstein relation
in terms of the {\em violation integral} %at NESS
\bea
I \equiv D_s - T\, \mu_s 
= \int_0^\infty d\t \left[  \la v(\t) \nu(0) \ra - \la v_s \ra^2 \right].
\label{Iv}
\eea
Since the form of velocity-MFDR for Langevin equation (Eq.~\ref{lange-vA}) 
and master equation (Eq.~\ref{master-vA}) are the same, we get the same 
modified Einstein relation for both the cases.

\section{Flashing ratchet model of molecular motors}
\label{ratchet}
In this section we apply the concepts developed so far in this paper
on a specific realization of the flashing ratchet model of molecular 
motors~\cite{Astumian1994,Prost1994,Julicher1997,Astumian2002,Kolomeisky2007}. 
In particular, we calculate the velocity-MFDR for this model and 
derive the violation integral of the corresponding non-equilibrium 
Einstein relation.

\label{ratchet}
\begin{figure}[t] % [htbp]
\begin{center}
\includegraphics[width=8 cm]{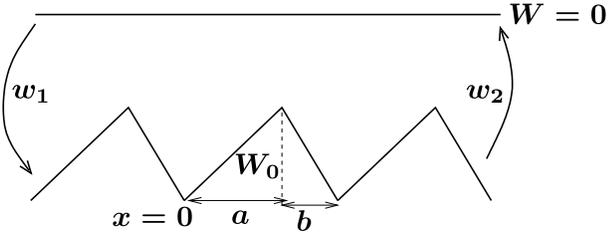}  
\caption{
Flashing ratchet model: The potential height switches between $W=0$ (off-state) and $W_0$ (on-state). 
The asymmetry of the potential in the on-state is described by the inequality $a\neq b$. 
$\w_{1,2}$ denote transition rates between on and off states. 
}
\label{model}
\end{center}
\end{figure}

A molecular motor, e.g., kinesin, moves along a polymeric track, e.g., microtubule in a 
strongly fluctuating thermal environment utilizing 
intrinsic local assymmetry of the track and chemical energy provided by hydrolysis of ATP to 
ADP and a phosphate. The binding and hydrolyzing of ATP changes the strength
of interaction of the motor with the polymeric track~\cite{Astumian1994}. 
Thus a simple two-state approximation of the dynamics of motor-proteins was proposed~\cite{Astumian1994,Prost1994} 
where the motor encounters a locally asymmetric
but globally periodic potential, whose height switches between a large and a small value.

We consider a flashing-ratchet model where the system swithces between two
states, (1)~on-state: stochastic motion in an asymmetric piece-wise linear 
potential, (2)~off-state:  simple one dimensional diffusion (Fig.~\ref{model}).
The probability distributions in the two states $p_{1,2}(x,t)$ evolve by~\cite{Prost1994}
\bea
\p_t p_1 + \p_x j_1 &=&  \w_2 p_2 - \w_1 p_1 \crcr
\p_t p_2 + \p_x j_2 &=& -\w_2 p_2 + \w_1 p_1 \nn
\eea
where $j_1 = -D \p_x p_1$  and $j_2 = -D [p_2 \p_x(W/T)+\p_x p_2]$, and $\w_{1,2}$ denote 
the transition rates. 
In the on-state, the potential $W(x)$ is periodic $W(x)=W(x+\l)$ with period $\l=(a+b)$.
Within one period, $W(x) = (W_0/a) x$ if $0\leq x<a$, and 
$W(x) = (W_0/b) (\l-x)$ if $a\leq x <\l$. %(see Fig.~\ref{model})

\begin{figure}[t] % [htbp]
\begin{center}
\includegraphics[width=8cm]{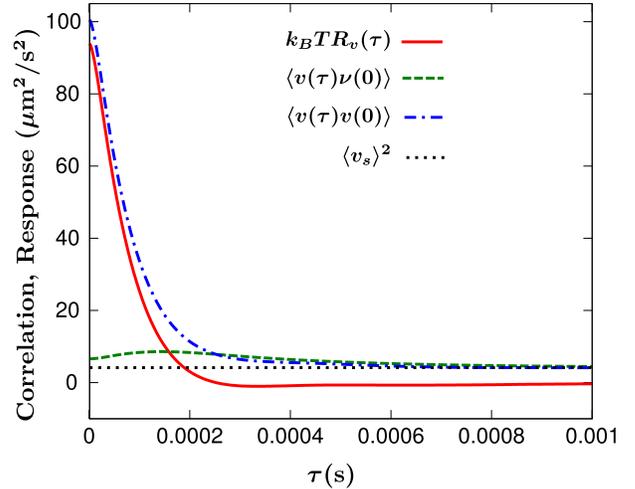} %{nu_corrln.eps} %{corrln.eps}  
\caption{(color online)
Velocity correlations and response: 
velocity response $\kb T\, R_v(\t)$,
correlation function of the velocity with the
local mean velocity $\la v(\t) \nu(0) \ra$, and 
velocity auto correlation function $\la v(\t) v(0) \ra$ as a function of time. 
The parameter-values used to obtain these curves are %\cite{Wang2003} 
$\lambda = 8\,$nm, $a/\l=0.1$,
$D = 0.009\,\mu$m$^2$s$^{-1}$, 
$k_BT = 4.2\,$pN~nm, $W_0 = 18.85\, \kb T$ and 
simulation time-step $\delta t = 1.8 \times 10^{-6}\,$s. The transition rates are chosen to be equal with 
$w_1 = w_2 = 3536\,$s$^{-1}$. With these parameter values we find
$D_s = 0.0084\,\mu$m$^2$s$^{-1}$, $\kb T\, \mu_s = 0.0057\,\mu$m$^2$s$^{-1}$, and 
the violation integral $I = 0.0027\,\mu$m$^2$s$^{-1}$.
The mean velocity in steady state is 
$\la v_s \ra=2.04\, \mu$m/s.
}
\label{corr}
\end{center}
\end{figure}

We perform molecular dynamics simulations of a particle moving under the influence of
the above-mentioned ratchet potential in the presence of a Langevin heat bath. 
We use stochastic switching between the on and off states with 
a constant switching rate $\w_1=\w_2$. 
From this simulation, in Fig.~\ref{corr}, we plot the
velocity-response function $\kb T R_v(\t)$ and the related 
steady-state correlations  $\la v(\t) v(0) \ra$,  $\la v(\t) \nu(0)\ra$ (Eq. \ref{fdt-v2}). 
The parameter values we use are enlisted in Fig.~\ref{corr} and 
are typical of microtubule associated molecular motors~\cite{Wang2003}. 
At long time, both $\la v(\t) v(0) \ra$ and $\la v(\t) \nu(0)\ra$ 
decorrelates to $\la v_s \ra^2$. 
We utilize the correlation functions to determine 
the mobility $\mu_s$, diffusion constant $D_s$, and the violation integral 
$I=D_s- \kb T\mu_s$ (see Fig.~\ref{corr}).

\begin{figure}[htbp]
\begin{center}
\includegraphics[width=8cm]{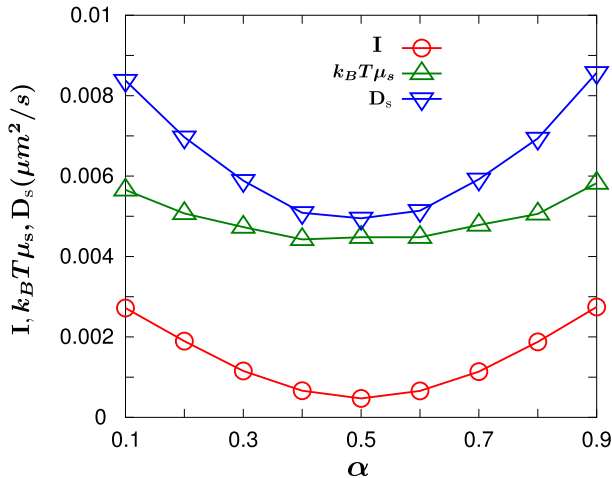}  
\caption{(color online) Flashing ratchet: diffusion constant $D_s$, mobility $\kb T \mu_s$ and violation integral  $I$
as a function of asymmetry parameter $\a=a/\l$. All the other parameter values are
same as in Fig~\ref{corr}. 
}
\label{alpha_I}
\end{center}
\end{figure}

We calculate the dependence of 
the steady-state mobility $\kb T\,\mu_s$, diffusion constant $D_s$ and the violation integral $I$ 
on the asymmetry parameter $\a=a/\l$ (Fig.~\ref{alpha_I}) where $\a=1/2$ denotes the symmetric 
ratchet. This calculation leads us to the curious result that all the three quantities have minimum at $\a=1/2$.  
The steady-state diffusion constant $D_s$ in the flashing-ratchet is always suppressed ($D_s<D$), 
and moves closer to the free diffusion $D$ for the most asymmetric ratchet. 
Note that the violation integral quantifies the difference between NESS and equilibrium,
with equilibrium requiring $I=0$. 
The symmetric ratchet does not generate unidirectional motion, but the switching 
between the on and off states keeps the system out of equilibrium. Thus, though 
the violation integral reaches its minimum at $\a=1/2$ it remains $I \neq 0$.
%{\textcolor{red}{
Setting switching rates $\w_1, \w_2 = 0$ would restore equilibrium with $I = 0$. %}} 
The dependence of $I$ on various models and parameter values
at different NESS is yet to be fully understood.

While calculation of all the other quantities from our simulations 
are straight-forward, $\nu(t)$ demands a special mention.
The local mean velocity $\nu(x)$ is the stochastic particle velocity $v$
averaged over the subset of trajectories passing through $x$.
At steady state, this definition is the same as $\nu_s(x)=j_s/p_s(x)$ where the 
mean current is constant everywhere: $j_s=\r \la v_s\ra $ with
$\la v_s \ra$ the mean velocity at steady state and $\r=1/\l$ the mean density.
In calculating $\la v(\t) \nu(0)\ra$, the local
mean velocity at time $t$ is obtained by identifying the value of 
$\nu(x)$ corresponding to the position $x$ visited by the particle 
at that instant.

 \section{Summary}
\label{summary}
We have presented a unified derivation of modified fluctuation dissipation relations 
(MFDR) at non-equilibrium steady states (NESS) using the Agarwal formalism. 
Thus all the various versions of MFDR that we derived in this paper are intrinsically
equivalent to each other. 
We showed that the response function around any NESS can be expressed as a correlation 
between the observable and a variable conjugate to the external force with respect to 
the system's stochastic entropy production.
For both a continuum Langevin and a discrete master equation system, 
we have shown that  the non-equilibrium form of FDT involving
velocity response can be expressed as an equilibrium one and an additive correction. The 
correction in both these cases is a correlation function of the velocity with 
a local mean velocity. The resulting modification of the Einstein's relation
gives the violation in terms of a time integral over this additive correction.

Using molecular dynamics simulations in presence of Langevin heat-bath, 
we studied a flashing ratchet model within this framework and obtained the 
response function and velocity correlations in the steady state. 
We showed that the violation integral varies non-monotonically with the asymmetry parameter of the 
ratchet and reaches a non-zero minimum for the case of a symmetric ratchet.
We plan to extend our study 
to other models of molecular motors~\cite{Kolomeisky2007}, %~\cite{Julicher1997,Korn2009,Kunwar2010}, 
stochastic particle-pumps~\cite{Jain2007, Chaudhuri2011}, 
polymer translocation dynamics~\cite{Muthukumar1999}, and dynamics of self-propelled particles~\cite{Vicsek2010}.

\acknowledgments
DC thanks Bela Mulder for useful comments and a critical reading of the manuscript.
The work of DC is part of the research program of the ``Stichting voor Fundamenteel 
Onderzoek der Materie (FOM)", which is financially supported by the ``Nederlandse 
organisatie voor Wetenschappelijk Onderzoek (NWO)". 
AC acknowledges support by grant EP/G062137/1 from the Engineering and Physical Sciences Research Council, UK.

\bibliographystyle{prsty}
%\bibliography{fdt}

\end{document}